\documentclass[%
 reprint,
superscriptaddress,
footinbib,
amsmath,amssymb,
prl,
nobibnotes
]{revtex4-1}

\usepackage[english]{babel}
\usepackage[utf8x]{inputenc}
\usepackage[colorinlistoftodos]{todonotes}
\usepackage{enumerate}
\usepackage{graphicx}
\usepackage{dcolumn}
\usepackage{bm}

\usepackage{textcomp}
\usepackage[pagewise,columnwise]{lineno}

\begin{document}

\preprint{APS/123-QED}


\title{
Nonextensivity and temperature fluctuations of the Higgs boson production
 }

\author{D. Rosales Herrera}
\affiliation{Facultad de Ciencias F\'isico Matem\'aticas, Benem\'erita Universidad Aut\'onoma de Puebla, Apartado Postal 165, 72000 Puebla, Pue., M\'exico}

\author{J. R. Alvarado García}
\email{j.ricardo.alvarado@cern.ch}
\affiliation{Facultad de Ciencias F\'isico Matem\'aticas, Benem\'erita Universidad Aut\'onoma de Puebla, Apartado Postal 165, 72000 Puebla, Pue., M\'exico}

\author{A. Fern\'andez T\'ellez}
\affiliation{Facultad de Ciencias F\'isico Matem\'aticas, Benem\'erita Universidad Aut\'onoma de Puebla, Apartado Postal 165, 72000 Puebla, Pue., M\'exico}

\author{J. E. Ram\'irez}
\email{jhony.eredi.ramirez.cancino@cern.ch}
\affiliation{Centro de Agroecología,
Instituto de Ciencias,
Benemérita Universidad Autónoma de Puebla, Apartado Postal 165, 72000 Puebla, Pue., M\'exico}

\author{C. Pajares}
\affiliation{
Departamento de F\'isica de Part\'iculas and Instituto Galego de Física de Altas Enerxías, Universidad de Santiago de Compostela, E-15782 Santiago de Compostela, Espa\~na}

\begin{abstract}

We determine the temperature fluctuations associated with the Higgs boson $p_T$ spectrum through the derivation of the string tension distribution corresponding to the QCD-based Hagedorn function, frequently used to fit the transverse momentum distribution (TMD). The identified string tension fluctuations are heavy tailed, behaving similarly to the $q$-Gaussian distribution. 
After the convolution with the Schwinger mechanism, both approaches correctly describe the entire TMD.
This approach is the onset for the nonthermal description of the particle production in ultrarelativistic pp collisions.
By analyzing the data of pp collisions at $\sqrt{s} =13$ TeV, we found that the average temperature associated with the Higgs boson differential cross section is around 85 times greater than the estimated value for the charged particle TMD.
Our results show that the Higgs boson production exhibits the largest deviation from the thermal description.

\end{abstract}
\maketitle


One fundamental problem in quantum chromodynamics (QCD) is understanding why quarks and gluons are never found in isolation but always forming color-neutral combinations (hadrons) \cite{Gell-Mann:1964ewy,Wilson:1974sk,Nambu:1976ay,McKellar:1986dr}.
Early models attempted to explain confinement through the formation of color flux tubes or color strings between partons. In these approaches, the potential energy in the strings increases linearly with the distance between partons, preventing their separation beyond a certain energy threshold \cite{Nambu:1974zg,Lizzi:1984wt,Morris:1987pt}.

In 1968, Veneziano proposed a framework for describing the scattering amplitudes of mesons \cite{Veneziano:1968yb}, which later led to the development of dual resonance models. These dual models exhibited properties reminiscent of strings, where particles emerged as resonances corresponding to different vibrational modes of the string \cite{Fubini:1969wp,Goto:1971ce,Mandelstam:1974fq,Schwarz:1973yz}. In the early 1970s, QCD was established as the fundamental gauge theory of the strong force describing the color interactions via gluon exchanges \cite{Gross:1973id,Politzer:1973fx,Fritzsch:1973pi,Gross:1974cs}. Shortly after, lattice QCD was proposed as an alternative approach to understanding the behavior of quarks and gluons at a fundamental level \cite{Wilson:1974sk}, confirming the confinement and providing insights into the nonperturbative aspects of QCD \cite{Luscher:2002pz}. 

The development of string-inspired models as effective descriptions of QCD acquired an important role in the low $p_T$ regime, where standard perturbative techniques are nonpractical. Phenomenological models such as the Lund string model \cite{andersson1998lund}, Color String Percolation Model \cite{Braun:2015eoa,bautista2019string}, or event generators, among which stands out PYTHIA (even EPOS and Herwig), take into account the string fragmentation to describe experimental observables as well as the production of particles in high-energy colliders (RHIC, LHC) \cite{Andersson:1983ia}. 

In particular, the study of the particle spectrum is useful to infer information about the properties of the produced medium, including the possible formation of the quark-gluon plasma. To this end, several fitting functions based on different assumptions have been proposed to describe the behavior of transverse momentum distribution (TMD).
For instance, Hagedorn introduced a QCD-based function that correctly describes the TMD behaviors, namely, exponential decay at low $p_T$ and a tail that decreases as a power-law at high $p_T$ \cite{Hagedorn:1983wk}. Later, Bialas considered the Schwinger mechanism together with the assumption that the string tensions fluctuate according to a Gaussian distribution \cite{BIALAS1999301}. This approach recovered the thermal $p_T$ spectrum, formerly obtained for hadron gas models.

Other efforts to describe the entire TMD data promote the exponential decay to a $q$-exponential function \cite{Wong:2015mba}. This approach is equivalent to the Hagedorn function \cite{Saraswat:2017kpg}, for which Wilk and W\l{}odarczyk derived 
from the convolution of the thermal distribution with temperature fluctuations \cite{Wilk:1999dr}.
This result connects the TMD observed with the local properties of the systems created in high energy collisions.
Nevertheless, descriptions of the string tension fluctuations have not yet been deduced for these approaches.

In recent studies \cite{Pajares:2022uts,Garcia:2022eqg}, it was found that the entire TMD of produced particles of pp collisions can be described by a family of Tricomi’s functions with two free parameters, obtained by considering the Schwinger mechanism and a $q$-Gaussian characterization of the string tension fluctuations, which are fine-tuned by analyzing the experimental TMD for each particular data set.
One immediate implication of these results is that the hard part of the TMD can be originated by raising the probability of observing strings with higher tension. In other words, the string tension fluctuations must be a heavy tailed distribution, giving rise to a nonthermal description of the systems.

In this paper, we aim to derive the string tension fluctuations that describe the QCD-based Hagedorn function as well as the temperature fluctuations corresponding to the $q$-Gaussian tension fluctuations. To do this, we derive the TMD from the convolution of the Schwinger mechanism with both temperature and tension fluctuations. 
To connect our findings with experimental results, we analyze the TMD data of charged particles produced in pp collisions 
under the minimum bias and V0 multiplicity classes reported by the ALICE Collaboration.  Moreover, we fit the TMD data of the Higgs boson produced in pp collisions at $\sqrt{s}$=13 TeV reported by CMS and ATLAS Collaborations in the combined channels $H\to \gamma \gamma$, $H\to ZZ^*$, and $H\to b\bar{b}$ (only CMS). 
The main result presented here is the determination and characterization of the temperature fluctuations associated with the TMD of produced particles in pp collisions, including the Higgs boson.

In string models, the production of particles is described through the creation of neutral color current pairs {$\text{q}$-$\bar{\text{q}}$} and {$\text{qq}$-$\bar{\text{q}}\bar{\text{q}}$} 
that later decay into the observed hadrons. In this way, the transverse momentum of these particles is given by the Schwinger mechanism
\begin{equation}
    \frac{dN}{dp_T^2} \sim e^{-\pi p_T^2/x^2},
\end{equation}
where $x^2$ is the string tension \cite{schwinger}. Here, we use the notation $dN/dp_T^2$ to indicate the invariant yield of produced particles. If the tension fluctuates according to a distribution $P(x)$, then the transverse momentum distribution must be computed as the convolution of the Schwinger mechanism with the string tension fluctuations. 
Thus
\begin{equation}
    \frac{dN}{dp_T^2} \propto \int_0^\infty e^{-\pi p_T^2/x^2} P(x) dx.
\end{equation}
In 1999, Bialas introduced a Gaussian description for the string tension fluctuations based on a stochastic QCD-vacuum approach,
given by \cite{BIALAS1999301}
\begin{equation}
P(x)= \sqrt{\frac{2}{\pi \varsigma^2}}e^{-x^2/2 \varsigma^2},
\label{eq:gaussian_flucts}
\end{equation}
with $\varsigma^2$ being the variance of the distribution.
After the convolution with the Schwinger mechanism, the TMD becomes a thermal distribution \cite{BIALAS1999301}
\begin{equation}
    \frac{dN}{dp_T^2} \propto e^{-p_T/T}, \label{eq:thermal}
\end{equation}
which resembles the Boltzmann distribution in the classical ensemble theory, where $T=\varsigma/\sqrt{2\pi}$ is identified as the \emph{temperature} of the TMD \cite{Braun-Munzinger:1994ewq,garcia2022percolation}.
From a statistical thermodynamic description, the final state could be characterized by an effective temperature $T$ associated with the medium wherein the particles are produced \cite{Hagedorn:1965st,Hagedorn:1967tlw}. This temperature-like parameter is also interpreted as the freeze-out temperature for particle spectra \cite{Ollitrault:2007du,Garcia:2022eqg}, suggesting that $T$ can be understood as a scale temperature of the medium created in the collision.

Equation~\eqref{eq:thermal} adequately describes the experimental TMD at low $p_T$ values. However, it considerably deviates from the TMD tail.
One reason for that is because of the possibility of hard gluon emissions from strings, as in the Lund model \cite{andersson1998lund}. Then, the production of high-momentum partons may give rise to the creation of massive particles \cite{Norrbin:2000zc,LHCb:2023wbo}. For instance, the principal channels of the Higgs boson production are the gluon fusion and the annihilation of vector bosons \cite{Georgi:1977gs,Cahn:1983ip,ATLAS:2022ooq}. 


\begin{figure*}[ht]
    \centering
    \includegraphics[scale=0.617]{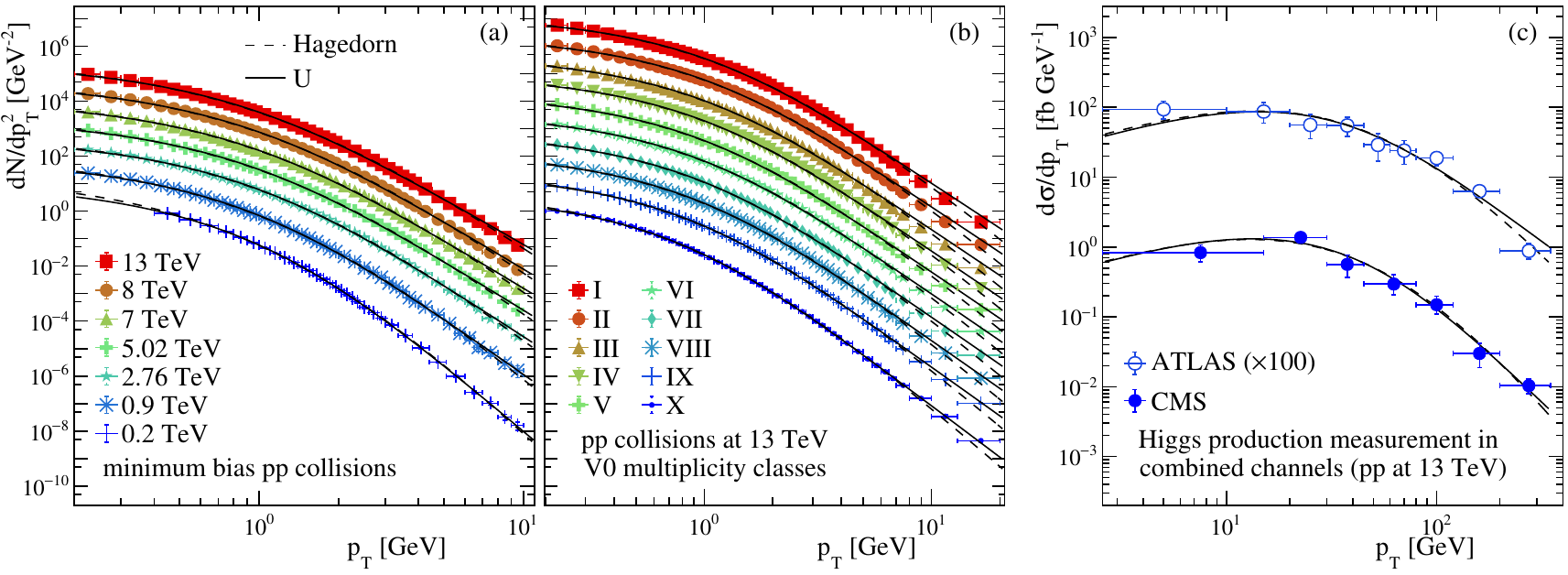}
\caption{Fits to experimental TMD data (figures) of charged particles for (a) minimum bias pp collisions at different center-of-mass energies, (b) ALICE-V0 multiplicity classes at $\sqrt{s} =$ 13 TeV, and (c) Higgs boson produced in pp collisions at $\sqrt{s} =$ 13 TeV. Solid and dashed lines correspond to $U$ and Hagedorn fitting functions, respectively. The data sets were scaled to improve visualization. }
    \label{fig:fits}
\end{figure*}

As we commented before, the TMD data is frequently described by the QCD-based function proposed by Hagedorn \cite{Hagedorn:1983wk}:
\begin{equation}
\frac{dN}{dp_T^2} \propto (1+p_T/p_0)^{-m}, \label{eq:Hag}
\end{equation}
where $p_0$ is a momentum scale threshold coming from the elastic scattering of the particles in the collision, and $m$ is a power law exponent \cite{Hagedorn:1965st,Hagedorn:1964zz,Hagedorn:1983wk}.
One advantage of the Hagedorn function is the correct description of the asymptotic behaviors of the TMD. At low $p_T$ values, $dN/dp_T^2\sim e^{-p_T/T_\text{Hag}}$ with $T_\text{Hag}=p_0/m$. At high $p_T$ values, the TMD decreases as a power law, $dN/dp_T^2\sim p_T^{-m}$ \cite{Hagedorn:1983wk}.
Equation~\eqref{eq:Hag} can be rewritten as the convolution of the thermal distribution~\eqref{eq:thermal} with temperature fluctuations \cite{Wilk:1999dr}, i.e., 
\begin{equation}
(1+p_T/p_0)^{-m}=\int_0^\infty e^{-p_T/T} \frac{\Gamma(1/T, m, p_0)}{T^2} dT,
\label{eq:HagfluctT}
\end{equation}
where $\Gamma(x, \alpha, \beta) = \beta^\alpha x^{\alpha-1}e^{-\beta x}  /\Gamma(\alpha)$ is the Gamma distribution.
Equation~\eqref{eq:HagfluctT} shows that the Hagedorn function emerges from the temperature fluctuations. 
We recall that the thermal distribution is obtained by considering that the tension fluctuates according to a Gaussian distribution~\eqref{eq:gaussian_flucts}.
Therefore, the Hagedorn function~\eqref{eq:Hag} is rewritten as follows
\begin{equation}
(1+p_T/p_0)^{-m}=\int_0^\infty \int_0^\infty e^{-\pi p_T^2/x^2} P(x) \mathcal{T}_\text{Hag}(T) dT dx,
\end{equation}
where $\mathcal{T}_\text{Hag}(T)=\Gamma(1/T, m, p_0)/T^2$ is the function describing the temperature fluctuations.
Notice that the integration of the joint probability $P(x)\mathcal{T}(T)$ with respect to $T$ gives the string tension fluctuations.
Therefore, for the Hagedorn function, we found
\begin{eqnarray}
P_\text{Hag}(x)&=&\int_0^\infty P(x) \mathcal{T}_\text{Hag}(T)dT \\
 & =& \frac{m p_0^m \pi^{\frac{m-1}{2}}  }{x^{m+1} }
    U\left(\frac{m+1}{2},\frac{1}{2},\frac{\pi p_0^2}{x^2}\right),
\label{eq:PHag}
\end{eqnarray}
where $U$ is the Tricomi's confluent hypergeometric function, defined as
\begin{equation}
U(a, b, z)=\frac{1}{\Gamma(a)}\int_0^\infty e^{-zt} t^{a-1}(1+t)^{b-a-1} dt. \label{eq:U}
\end{equation}
Thus,
\begin{equation}
(1+p_T/p_0)^{-m}=\int_0^\infty e^{-\pi p_T^2/x^2} P_\text{Hag}(x) dx.
\label{eq:Hagtension}
\end{equation}
Equation~\eqref{eq:Hagtension} means that the Hagedorn function is also obtained from the picture of the formation of color string clusters where their tension fluctuates according to $P_\text{Hag}(x)$ in Eq.~\eqref{eq:PHag}.
Let us discuss the asymptotic behaviors of $P_\text{Hag}$. 
At low $x$ values, 
\begin{equation}
P_\text{Hag} \propto 1-\frac{(m+1)(m+2)x^2}{4\pi p_0^2}+\mathcal{O}(x^4) \sim e^{-x^2/2\beta_\text{Hag}^2}.
\end{equation}
Hence, $P_\text{Hag}$ recovers the Gaussian string tension fluctuations with variance $\beta_\text{Hag}^2=2\pi p_0^2/[(m+1)(m+2)]$.
On the other hand, at high $x$ values, we found $P_\text{Hag}\sim x^{-(m+1)}$.
This implies that $P_\text{Hag}$ is a heavy-tailed distribution.

By plugging $p_0=T_\text{Hag} m$ in Eq.~\eqref{eq:Hag}, the Hagedorn function becomes a $q$-exponential Tsallis function. Then, the TMD is rewritten as $dN/dp_T^2 \propto (1+p_T/m T_\text{Hag})^{-m}$.
By taking $m=1/(q_e-1)$ or $m=q_e'/(q_e'-1)$, the Hagedorn function recaptures different versions of the $q$-exponential Tsallis distribution frequently used by the high energy physics community to fit the TMD experimental data \cite{Bialas:2015pla,Deb:2019yjo,biro2020tsallis}.


Another interesting heavy tailed distribution frequently used in the literature to study nonextensive and nonequilibrium phenomena is the $q$-Gaussian Tsallis function, given by \cite{budini2}
\begin{equation}
P_{q\text{G}}(x) = \frac{2 z_0^{\frac{1}{2}} \Gamma\left( \frac{1}{q-1} \right)} { \pi \Gamma\left( \frac{1}{q-1} - \frac{1}{2}\right)}
    \left( 1 + \frac{z_0 x^2}{\pi}\right)^{\frac{1}{1-q}}
    \label{eq:qGfluc}
\end{equation}
where $z_0 = \pi(q-1)/2\sigma^2$, $q$ is a deformation parameter \footnote{$q$ must be bounded by 1 and 3/2 to assure that $\langle p_T  \rangle$ is finite.}, and $\sqrt{\sigma^2}$ is the width of the distribution.
$P_{q\text{G}}$ has the following asymptotic behaviors: at low $x$ values,
\begin{equation}
P_{q\text{G}}\propto 
1
-\frac{x^2}{2 \sigma^2} 
+\mathcal{O} (x^4) \sim e^{-x^2/2\beta_U^2},
\end{equation}
where $\beta_U^2= \sigma^2$ is identified as the variance of the Gaussian fluctuations recovered in this limit.
Conversely, the tail of $P_{q\text{G}}$ behaves as $(x^2)^{-1/(1-q)}$.
The convolution of the Schwinger mechanism with $P_{q\text{G}}$ results in a confluent hypergeometric function \eqref{eq:U}. 
In this way, the TMD is described by 
\begin{equation}
\frac{dN}{dp_T^2} \sim U\left(\frac{1}{q-1}-\frac{1}{2},\frac{1}{2},z_0 p_T^2\right). \label{eq:TMDU}
\end{equation}
The latter has the appropriate asymptotic behaviors. At low $p_T$ values, $dN/dp_T^2\sim \exp(-p_T/T_U)$, with thermal temperature $ T_U=  B(a , 1/2)/\sqrt{4\pi z_0}$, where $B$ is the Beta function. 
On the other hand, the TMD behaves as $(p_T^2)^{\frac{1}{2}-\frac{1}{q-1}}$ for high $p_T$ values \cite{Pajares:2022uts,Garcia:2022eqg}. 

\begin{figure}[ht]
    \centering
\includegraphics[scale=0.442]{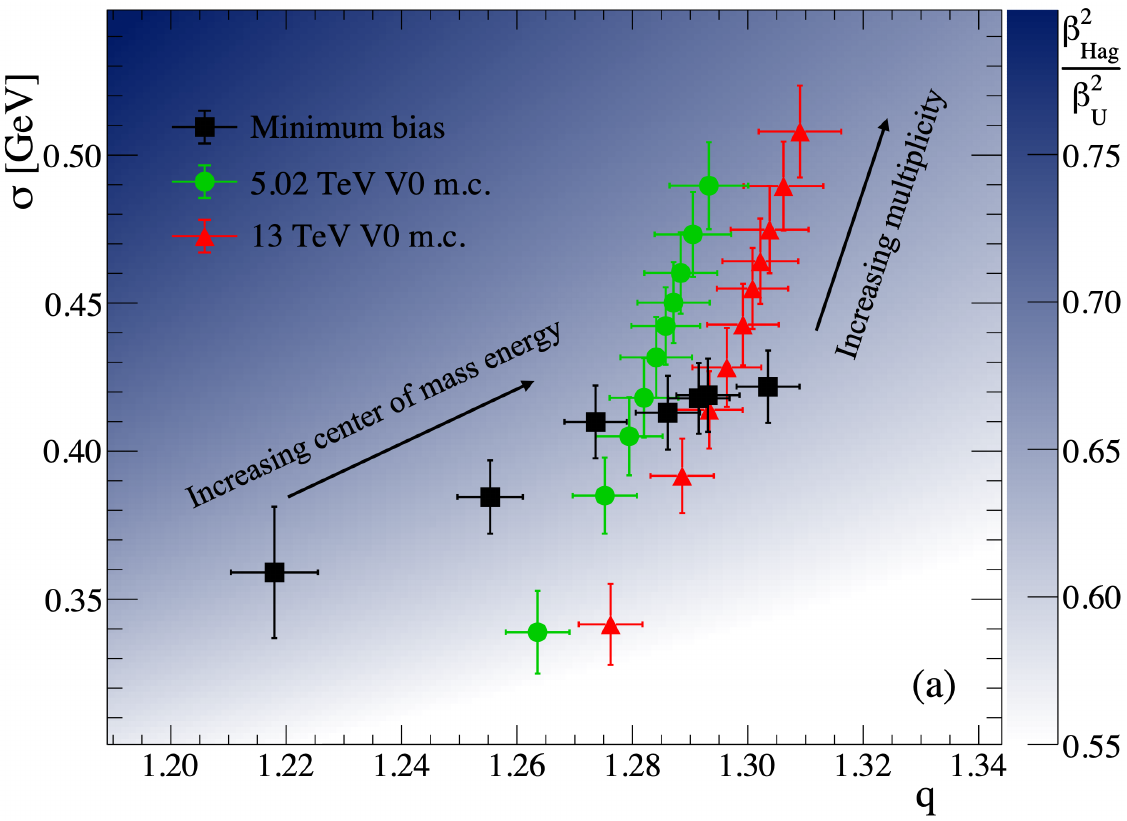}
\includegraphics[scale=0.294]{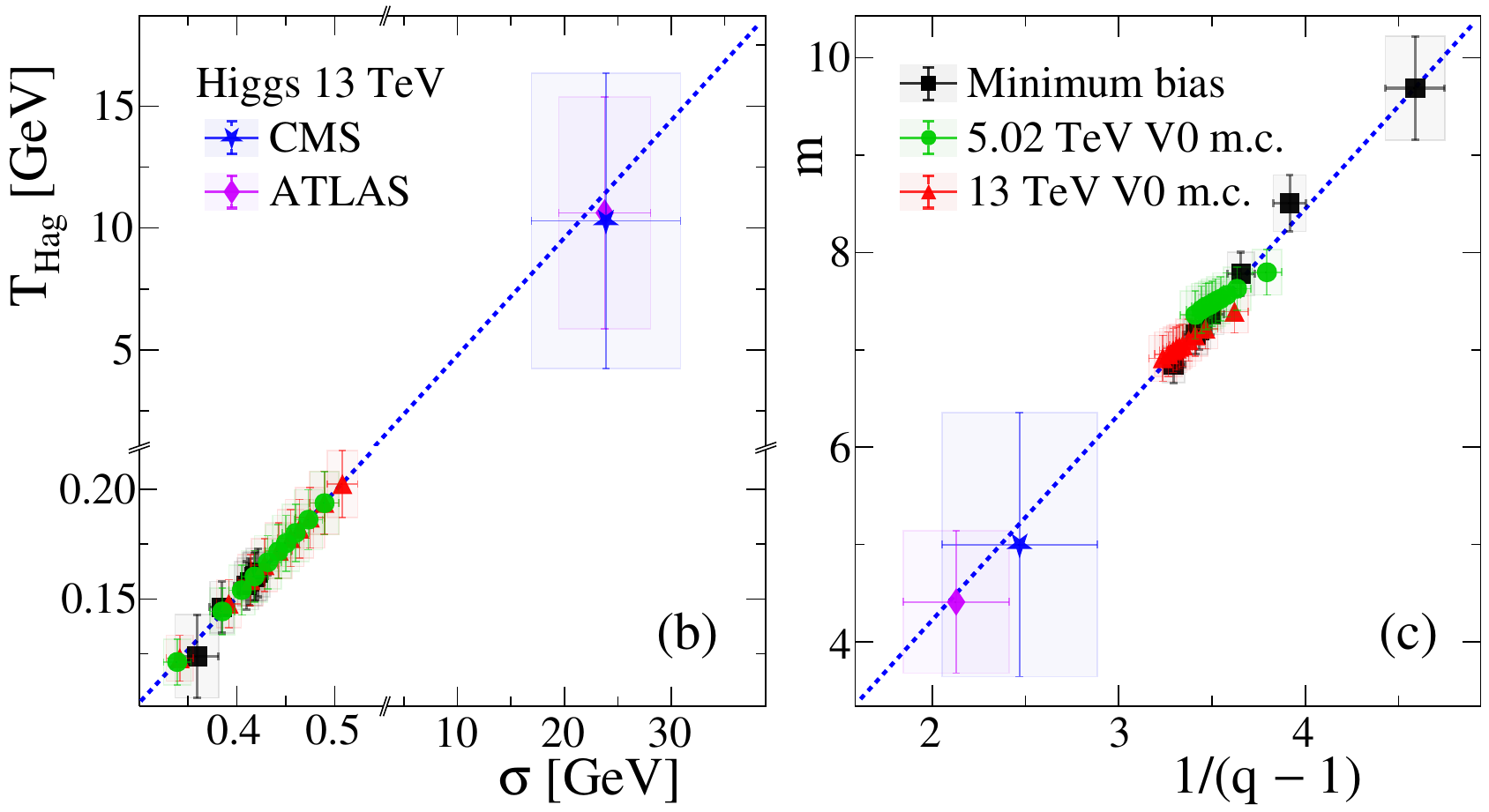}
\caption{(a) Ratio between the variance $\beta^2_\text{Hag}/\beta^2_U$ of the Hagedorn and $U$ functions (gradient color) in the $q$-$\sigma$ parameters space. 
Figures correspond to the pp collisions 
experimental data analyzed.  Correlations between the fitting parameters of the $U$ and Hagedorn functions for the pp collisions data analyzed: (b) $T_\text{Hag}$ vs $\sigma$ and (c) $m$ vs $1/(q-1)$. Dotted lines correspond to the linear trends.}
    \label{fig:params}
\end{figure}

Similarly to the Hagedorn case, the $q$-Gaussian distribution can be expressed as the convolution of the Gaussian string tension fluctuations \eqref{eq:gaussian_flucts} with the following temperature fluctuations
\begin{equation}
     \mathcal{T}_U(T)
     =
  \frac{2}{T^3} \Gamma\left( \frac{1}{T^2},
 \frac{1}{q-1} - \frac{1}{2}, \frac{1}{4z_0}    \right).
\end{equation}
Hence, the TMD~\eqref{eq:TMDU} is rewritten as follows
\begin{eqnarray}
\frac{dN}{dp_T^2} 
&\sim& 
      \int_0^\infty \int_0^\infty e^{-\pi p_T^2/{x^2}} P(x)  \mathcal{T}_U(T)   dT dx   
          \nonumber
    \\ &=&
      \int_0^\infty e^{- p_T/{T}} \mathcal{T}_U(T)   dT   .  
      \label{eq:UfluctT}
\end{eqnarray}
Again, the TMD is expressed as the convolution of the thermal distribution with the temperature fluctuations. Note that Eqs.~\eqref{eq:HagfluctT} and \eqref{eq:UfluctT} can be identified as Laplace-like transforms of $\mathcal{T}_\text{Hag}$ and $\mathcal{T}_U$, respectively.
We must emphasize that both $P_\text{Hag}$ and $P_{q\text{G}}$ are heavy-tailed distributions, which characterize the departure from the thermal description~\eqref{eq:thermal}. Moreover, the production of high-$p_T$ particles (including massive particles and jets) is implicitly incorporated via the observations of rare events modeled by the tails of these distributions.

\begin{figure}[ht]
    \centering
\includegraphics[scale=0.294]{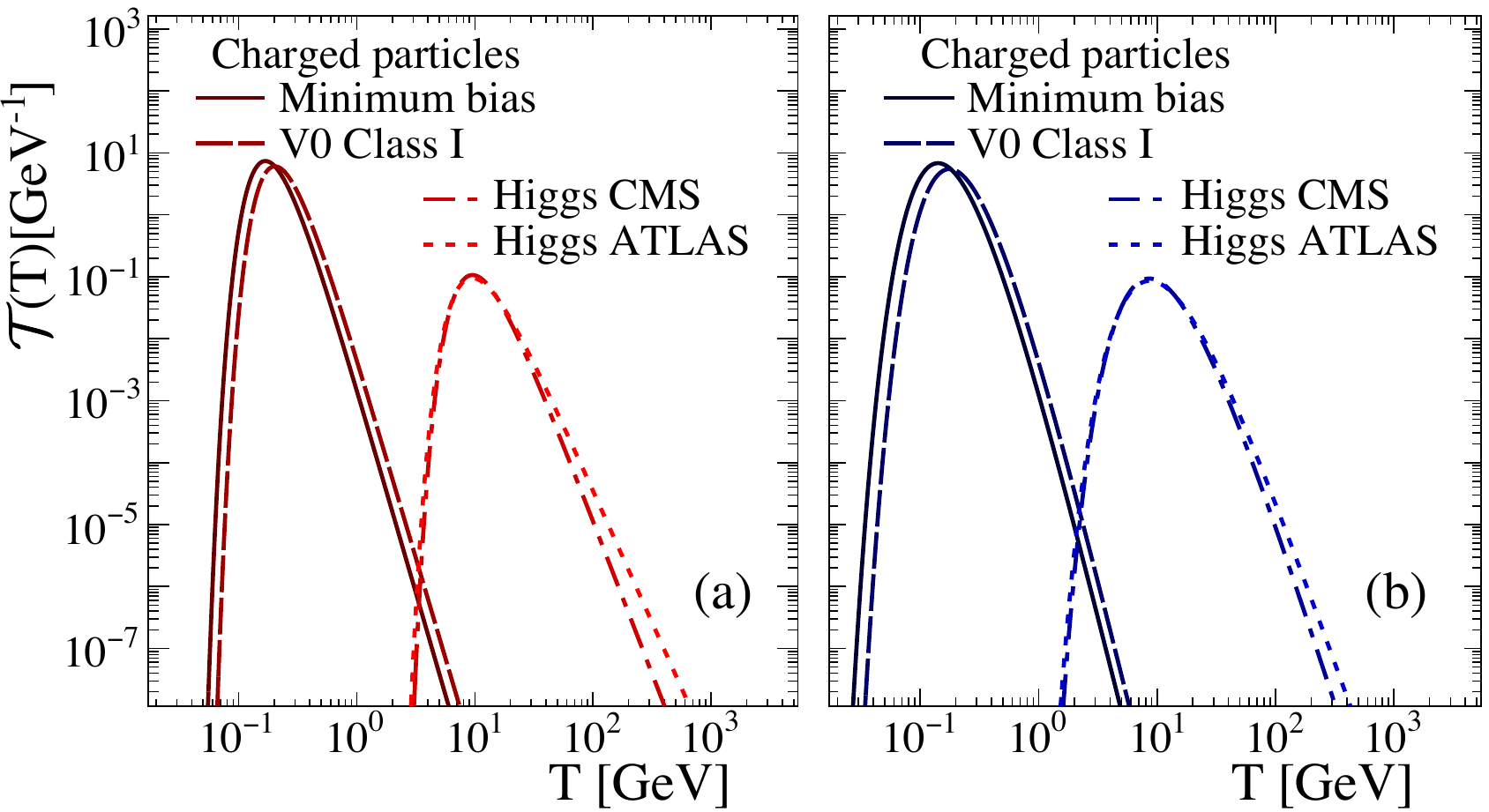}
\includegraphics[scale=0.294]{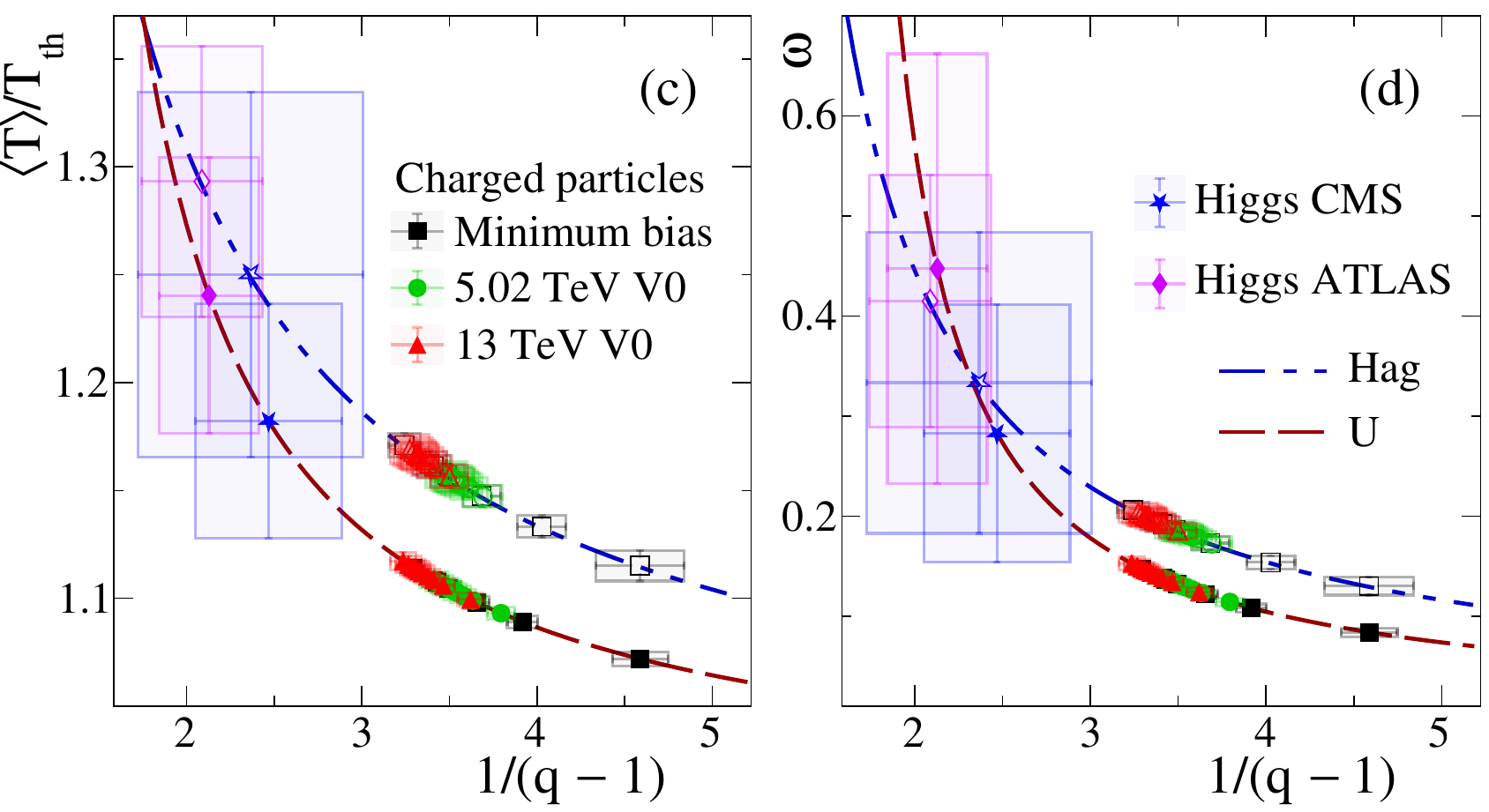}
\caption{
temperature fluctuations (a) $\mathcal{T}_U$ and (b) $\mathcal{T}_\text{Hag}$ obtained for charged particle production in pp collisions at $\sqrt{s} = $ 13 TeV under minimum bias conditions (solid lines) and processed with the ALICE V0 classes (dashed lines). Dotted and dash-dotted lines correspond to the temperature fluctuations associated with the Higgs boson TMD obtained by analyzing the ATLAS and CMS data, respectively.
(c) $\langle T\rangle/T_\text{th}$ and (d) $\omega$ as a function of $1/(q-1)$ for the $\mathcal{T}_U$ (filled figures) and $\mathcal{T_\text{Hag}}$ (empty figures). The dashed and dash-dotted lines are the theoretical determinations \eqref{eq:meanT} and \eqref{eq:varT}.}
    \label{fig:fluct}
\end{figure}

In both cases, $\mathcal{T}_\text{Hag}$ and $\mathcal{T}_U$ correspond to local temperature fluctuations wherein a particle is produced. In consequence, the temperature variations occur in small parts of the system, implying that the system is nonthermal.

  
The particular form of the string tension and temperature fluctuations depend on the free parameters of~\eqref{eq:Hag} and~\eqref{eq:TMDU}, which are determined by fitting different data sets. To this end, 
we analyze the experimental TMD data of the charged particle production reported by the STAR and ALICE Collaborations for pp collisions considering different classes of events, namely, minimum bias \cite{STAR:2003fka,ALICE:2010syw,ALICE:2022xip} at different center-of-mass energies and the data reported by the ALICE Collaboration using the V0 multiplicity classes \cite{ALICE:2019dfi}. 
We also analyze the data of the differential cross section ($d\sigma/dp_T$) of the Higgs boson reported by the ATLAS and CMS Collaborations in the combined channels $H\to \gamma \gamma$, $H\to ZZ^{*}$, and $H\to b\bar{b}$ (only CMS) \cite{ATLAS:2018pgp,CMS:2018gwt}.

The data is processed by using the ROOT 6 software.    
In Fig.~\ref{fig:fits}, we show the experimental data together with their corresponding fits according to the Hagedorn \eqref{eq:Hag} and Tricomi \eqref{eq:TMDU} fitting functions. Notably, both approaches accurately describe the TMD data (in all cases $\chi^2/\text{ndf} < 1$).

For the analyzed data, we found that the fitting parameters of the two approaches can be linearly related, as shown in Fig.~\ref{fig:params}.
The parameters satisfy the relations: $m=a_q/(q-1)$, and $T_\text{Hag}=p_0/m=m_\sigma \sigma +b_\sigma$, with $a_q$=2.112(7), $m_\sigma$=0.482(5), and $b_\sigma$=-0.042(2) GeV.
These relations are relevant because they establish a mapping between the Hagedorn and $q$-Gaussian approaches.
Hence, we can describe all the observables as a function of two parameters, for instance, $q$ and $\sigma$.

Interestingly, for all the cases analyzed, we found that $T_\text{Hag}<T_U$.
This observation can be tracked from the string tension fluctuations of each approach. We recall that analyzing the TMD behavior at low $p_T$ defines the soft scale. 
Notice that both $P_\text{Hag}$ and $P_{q\text{G}}$ tend to a Gaussian distribution in the limit of low $x$ values.
Then, the corresponding thermal picture of these limits leads to soft scales proportional to the variance's square root.
We found $\beta_\text{Hag}< \beta_U$ for the analyzed TMD data. Figure~\ref{fig:params} (a) shows the ratio $\beta_\text{Hag}^2/ \beta_U^2$, which is less than 1 in all cases.
This implies that the $P_\text{Hag}$ is a more narrow distribution than $P_{q\text{G}}$. In fact, $P_\text{Hag}$ substantially increases the probability of observing strings with tension around zero compared with the Gaussian and $q$-Gaussian distributions.
In consequence, $P_\text{Hag}$ produces an excess of strings with tension close to zero. Hence, the production of particles with very low $p_T$ is enhanced but in a short $p_T$ range. The main implication is that the soft scale ($T_\text{Hag}$) is underestimated. This is expected because the Hagedorn function was proposed in order to explain the TMD produced by hard QCD processes.

Let us discuss some statistics of the temperature distribution. The average temperature is
\begin{subequations}
\begin{eqnarray}
   \langle T  \rangle_{U} &=& \frac{(q-1)B\left(\frac{1}{q-1}, -\frac{1}{2} \right)^2}{4\pi (2-q)} T_U,\\
   \langle T  \rangle_\text{Hag} &=& \frac{m}{m-1} T_\text{Hag}.
\end{eqnarray}
    \label{eq:meanT}
\end{subequations}
In both cases, the average temperature is greater than the soft scale as the system becomes nonthermal, but they coincide only in the thermal limit $q\to 1$ and $m\to \infty$ for the $U$ and Hagedorn functions. On the other hand, when the parameters $q$ and $m$ are close to their limit values ($q<1.5$ and $m>3$ to assure the convergence of $\langle p_T \rangle$), the quotient $\langle T  \rangle/T_{th}$ goes to $4/\pi$ and 3/2 for the $U$ and Hagedorn functions, respectively. In Fig.~\ref{fig:fluct} (c), we show the values of $\langle T  \rangle/T_{th}$ for the data analyzed under the $U$ and Hagedorn fitting functions.
Notice that both the soft scale and the average temperature grow as the center-of-mass energy and multiplicity do. In particular, the Higgs boson spectra have a temperature one hundred times greater than any analyzed TMD of charged particles. It indicates that the soft scale may also depend on the particle masses.
We must point out that the Higgs boson is not produced through soft processes. However, its $p_T$-spectrum encompasses around 54\% of the produced Higgs bosons with $p_T$ lower than 30 GeV, which can be described by a $p_T$-exponential. This range of $p_T$ can be considered the soft part of the TMD.
It is worth mentioning that, in both cases, the average of $1/T$ takes the values of the inverse of the soft scales, such as Wilk and W\l{}odarczyk discussed in \cite{Wilk:1999dr}.

We also determine the relative variance of the temperature distributions as $\omega = \text{var(T)}/ \langle T  \rangle^2$, obtaining closed formulae depending only on the parameters that modulate the shape of the TMD tail: 
\begin{subequations}
\begin{eqnarray}
    \omega_{\text{Hag}} &=& \frac{1}{m-2}, 
    \\
    \omega_U &=& \frac{8\pi (q-2)^2}{(q-1)(5-3q) B\left(\frac{1}{q-1}, -\frac{1}{2} \right)^2 } -1.
\end{eqnarray}\label{eq:varT}
\end{subequations}
Interestingly, $\omega$ diminishes as the system reaches the thermal behavior. In this limit, the temperature fluctuations are described by the Dirac's delta $\delta(T - T_{th})$. 
This means that the temperature is the same along with the medium wherein the particles are produced. However, as the $p_T$ spectrum becomes nonthermal, the particle production may occur at different temperatures according to the distributions $\mathcal{T}_U$ and $\mathcal{T}_{\text{Hag}}$. 

In Fig.~\ref{fig:fluct} (d), we show the $\omega$ dependence on $1/(q-1)$ for the analyzed data using the $U$ and Hagedorn parameters, showing a monotonic increasing dependence on the nonextensivity parameters. Additionally, in Ref. \cite{garcia2022percolation}, the authors found a similar behavior for the relative variance of the string tension fluctuation. This means that larger fluctuations of the string tension imply larger fluctuations in the temperature. 
Similarly to $\langle T \rangle$, the relative variance increases with the center-of-mass energy and multiplicity for the production of charged particles in pp collisions, making the temperature fluctuations wider. 
In particular, for the case of the Higgs boson production, we found that the TMD has a large soft scale ($T \sim$ 11 GeV) and exhibits the largest possible deviation from the thermal description ($q \sim$ 1.45), which makes the temperature fluctuations wider with a large average temperature. 

In summary, we have proved that the QCD-based Hagedorn function can be derived from string tension fluctuations. Its distribution $P_\text{Hag}$ has similar asymptotic behaviors as the recently introduced $q$-Gaussian distribution. They recover a Gaussian functional form at low $x$ but both are power laws at high tension values.
The heavy tail of $P_{q\text{G}}$ and $P_\text{Hag}$ lead the production of particles with high $p_T$, marking the departure from the thermal distribution.
Thus, the involved physics in the Hagedorn function becomes compatible with color string models.

Using the derived functions, we analyzed the experimental TMD data of pp collisions under different classes of events, for instance, minimum bias as a function of the center-of-mass energy, and V0 multiplicity classes.  
For each analyzed data set, we found the free parameters of the Tricomi's and Hagedorn functions, which determine the string tension and temperature fluctuations and reveal specific conditions for the particle production.

Note that the Hagedorn \eqref{eq:Hag} and the Tricomi \eqref{eq:TMDU} functions describe the entire spectrum. 
However, the Hagedorn function underestimates the soft scale because $P_{\text{Hag}}$ is more narrow than the $q$-Gaussian distribution.

Interestingly, the string tension fluctuations $P_\text{Hag}$ and $P_{q\text{G}}$ arise from the convolution of the Gaussian string tension with fluctuations in the temperature, which can be identified as the TMD in the temperature space.
For the temperature fluctuations, we found that the average temperature associated with the Higgs boson differential cross section is 85 times greater than the estimated for the charged particle TMD of minimum bias pp collisions at $\sqrt{s}=13$ TeV. This means that the system has a region with a very high temperature, which causes the flattening of the invariant yield's soft part and extends the range of $p_T$ where it is described by a thermal distribution. This characterization may correspond to the possibility of the creation of \emph{hot spots} when massive particles are produced.

In the color string percolation model, more than two strings may overlap, indicating that they interact through the color field. 
Following the color sum rules, this overlapping enhances the resultant tension of the cluster, being maximum for fully overlapped strings \cite{bautista2019string}. 
For this approach, it was found that, for the hadron production, the string density grows with the temperature parameter \cite{ramirez2021interacting}. Assuming this trend is still valid for heavier particles, the processes with the highest temperatures correspond to the highest string densities, as in the case of the Higgs boson production. One way to obtain the necessary energy density to create at least a parton required for massive or high-$p_T$ particle production comes from the possibility of the accumulation of strings in a singular spot \cite{garcia2022percolation}. Additionally, the heavy tailed distribution for the string tension increases the probability of having strings with large tensions, which may help to reach such energy density \cite{Pajares:2022uts}. In this sense, hot spots can be interpreted as places with high string density, i.e., a very dense string cluster, which corresponds to a region with a high local temperature. 
It is following the recent discussion about the production of jets with higher energy than expected by the Feynman scaling. This phenomenon extends beyond heavy nuclei and also encompasses smaller scale events. Therefore, the production of massive particles is linked to high-density spatial points that can be modeled through QCD string interactions \cite{Bland:2021xhg}. 


We must clarify that the Higgs boson production (or other massive particles) does not occur through thermal processes. 
However, the approaches discussed in this manuscript implicitly involve the production of high $p_T$ partons through the heavy tail of the string tension fluctuations. 
In this way $P_\text{Hag}$ and $P_{q\text{G}}$ model nonextensive and nonequilibrium systems. 
Moreover, the temperature fluctuations are a consequence of the deviation of the TDM from the thermal behavior. Since the $q$ parameter quantifies the nonextensivity of the systems, we found that the Higgs boson TMD exhibits the largest possible deviation from the thermal behavior.

Finally, our results can be useful in improving the string based computational tools used to simulate high energy collisions.

\begin{acknowledgments}
This work has been funded by the projects PID2020-119632GB-100 of the Spanish Research Agency, Centro Singular de Galicia 2019-2022 of Xunta de Galicia and the ERDF of the European Union.
This work was funded by Consejo Nacional de Humanidades, Ciencias y Tecnologías (CONAHCYT-México) under the project CF-2019/2042,
graduated fellowship grant number 1140160, and postdoctoral fellowship grant numbers 289198 and 645654.
J. R. A. G. acknowledges financial support from Vicerrectoría de Investigación y Estudios de Posgrado (VIEP-BUAP).
We also thank N. Armesto, A. Raya, and E. Cuautle for their valuable comments.
\end{acknowledgments}

\bibliography{ref}

\end{document}